\documentclass[final,1p,times]{elsarticle} 

\usepackage{graphicx}
\usepackage{amssymb} 
\usepackage{amsthm} 
\usepackage{lineno}
\journal{Nuclear Physics A} 

\begin{document}

\begin{frontmatter} 

\title{Measurements of the Correlation between Reconstructed Jets and the Reaction Plane in STAR at RHIC}

\author{Alice Ohlson (for the STAR\fnref{col1} Collaboration)}
\fntext[col1] {A list of members of the STAR Collaboration and acknowledgements can be found at the end of this issue.}
\address{Yale University, New Haven, CT, USA}

\begin{abstract} 
The relationship between jet properties and the underlying geometry of the medium produced in heavy ion collisions can be explored through a measurement of the correlation between the axes of reconstructed jets and the reaction plane, defined as ``jet $v_2$.''  Such a measurement provides information on the pathlength dependence of medium-induced parton energy loss and may also be used to assess biases in jet-finding methods.  

We present first measurements of jet $v_2$ in $\sqrt{s_{NN}}$ = 200 GeV Au+Au collisions in the STAR experiment at RHIC.  In order to reduce the artificial jet -- event plane bias, which results from jet fragments being included in the event plane calculation, detectors at forward pseudorapidity are used to determine the event plane when measuring the $v_2$ of reconstructed jets at mid-rapidity.  These measurements demonstrate a non-zero jet $v_2$, which is indicative of pathlength-dependent parton energy loss.  
\end{abstract} 

\end{frontmatter} 


\section{Introduction to Jet $v_2$}
The aim of jet physics in ultrarelativistic heavy ion collisions is to investigate parton energy loss in the strongly-coupled quark-gluon plasma produced in such collisions.  The energy loss, or quenching, of high-$p_T$ hadrons and jets has been demonstrated at RHIC and the LHC \cite{STARraa, PHENIXraa, ALICEraa, CMSraa, STARdihadron, PHENIXdihadron, ATLASdijet}.  Theoretical models indicate that parton energy loss depends on the length of the in-medium path that the parton traverses \cite{radcoll}.  It is also expected that the pathlength depends on the relative angle between the parton emission angle and the reaction plane, such that the pathlength is on-average shorter when the parton is emitted in-plane than when the parton is emitted out-of-plane \cite{xinnian}. 

While the emission angles of partons produced in hard scatterings should be independent of the initial QGP geometry (in the plane transverse to the beam direction), pathlength-dependent jet suppression can give rise to a difference in the number of jets reconstructed parallel and perpendicular to the event plane, depending on the biases involved in the jet reconstruction (introduced by $p_T$ cuts or the resolution parameter $R$).  This effect would result in a correlation between reconstructed jets and the reaction plane (or second-order participant plane), defined as ``jet $v_2$,'' which is defined quantitatively in Equation~\ref{eq:jetv2},
\begin{equation}\label{eq:jetv2}
v_2^{jet} = \frac{\left\langle \cos\left(2(\phi_{jet}-\Psi_{EP})\right)\right\rangle}{Res}
\end{equation}
where $\phi_{jet}$ is the azimuthal angle of the reconstructed jet axis, $\Psi_{EP}$ is the azimuthal angle of the event plane, and $Res$ is the event plane resolution.  

Jet $v_2$ is not synonymous with ``jet flow,'' since it does not have a hydrodynamic interpretation, but rather it is an indication of pathlength-dependent jet quenching.  An experimental constraint on the pathlength dependence of jet quenching may be able to differentiate between theories of medium-induced energy loss.  Since jets experiencing varying amounts of medium-induced modification can be selected with different jet reconstruction parameters, a measurement of jet $v_2$ can also lead to insights about the biases involved in jet finding.

\section{Jet Reconstruction in STAR} \label{sec:jet}
The data analyzed here were collected in 2007 in $\sqrt{s_{NN}} = 200$ GeV Au+Au collisions in the STAR detector at RHIC.  Events containing high-$p_T$ processes were selected by an online high tower (HT) trigger, which required $E_{T} > 5.4$ GeV deposited within a single tower (angular size $\Delta\phi \times \Delta\eta = 0.05 \times 0.05$) of the Barrel Electromagnetic Calorimeter (BEMC).  An offline cut selected events in which the HT trigger tower had $E_{T} > 5.5$ GeV after the charged energy deposition in the calorimeter was accounted for (by subtracting the momentum of any charged track pointing to a tower from that tower's energy).  

In this analysis, jets are reconstructed with the anti-$k_{T}$ algorithm from the FastJet package \cite{fastjet} with a resolution parameter of $R = 0.4$.  Only charged tracks in the Time Projection Chamber (TPC) with $p_T > 2$ GeV/$c$ and neutral towers in the BEMC with $E_T > 2$ GeV are used in the jet reconstruction, and the jets are required to contain the HT trigger tower.  The combination of the constituent $p_T$ cut and the 5.5 GeV HT trigger selects a jet sample which is highly-biased towards hard-fragmenting jets, which are more likely to be unmodified by the medium, potentially due to surface bias \cite{Renk_Dihadron}.  

\subsection{Background Fluctuations and the Jet Energy Scale}
Although the constituent $p_T$ cut reduces the effects of background fluctuations on the jet energy scale, it is still necessary to assess the effects of non-jet particles being clustered into the jet on the measurement of jet $v_2$.  Background particles (with $p_T > 2$ GeV/$c$) have significant $v_2$ \cite{STARv2} and are therefore more likely to be clustered into the jet cone in-plane versus out-of-plane, since the $v_2$ modulation of the background is not accounted for in the jet reconstruction.  Consequently, more low-$p_T$ jets get reconstructed at higher $p_T$, artificially increasing the number of in-plane jets in a fixed reconstructed jet $p_T$ range.  In this way, background fluctuations produce an artificial jet $v_2$ signal.  

The magnitude of this artificial jet $v_2$ is determined by embedding $p$+$p$ HT jets in Au+Au minimum bias events.  In this embedding, it is possible to access three relevant quantities: the reconstructed jet $p_T$ in $p$+$p$ ($p_{T}^{jet, p+p}$), the reconstructed jet $p_T$ with Au+Au background fluctuations  ($p_{T}^{jet, emb}$), and the event plane of the Au+Au event (determined before embedding the jet).  The jets are embedded isotropically with respect to the event plane of the underlying event, and therefore $v_2^{jet}$ is zero when calculated in a given range of $p_{T}^{jet, p+p}$.  However, when jet $v_2$ is calculated as a function of $p_{T}^{jet, emb}$, background fluctuations produce a measured artificial jet $v_2$ of approximately 4\% (with only small dependencies on centrality and $p_{T}^{jet}$).  In this analysis, the effect of background fluctuations is subtracted from the measured $v_2$ values directly, and we do not attempt to correct the reconstructed jet energy via an average background subtraction or any other method.

\section{Event Plane Determination}
When jet fragments are included in the calculation of the event plane, the event plane is preferentially reconstructed aligned with the jet axis.  Simple simulations show that this jet -- event plane bias can result in a 10-20\% artificial jet $v_2$.  This bias can be reduced or avoided by introducing a pseudorapidity ($\eta$) gap between the jet (and the recoil jet) and the particles used to calculate the event plane.  

The forward capabilities of STAR are used to determine the event plane in three $\eta$ ranges: the TPC covers midrapidity ($|\eta| < 1$), while the Forward Time Projection Chambers (FTPC) measure charged tracks in $2.8 < |\eta| < 3.7$, and Zero Degree Calorimeter Shower Maximum Detectors (ZDC-SMD) measure the energy deposition of spectator neutrons with $|\eta| > 6.3$.  Since the axes of the reconstructed jets are restricted to within $|\eta| < 0.6$, the FTPC and ZDC-SMD allow $\eta$ gaps of $|\Delta\eta| > 2.2$ and $|\Delta\eta| > 5.7$, respectively.  Furthermore, PYTHIA indicates that it is highly unlikely for the recoil jet axis to be within the pseudorapidity coverage of the FTPC (occurring fewer than ten times in two million PYTHIA events with $\hat{p}_T > 10$ GeV/$c$).  In the TPC and FTPC, which are track-based detectors, the event plane was calculated according to Equation~\ref{eq:EP},
\begin{equation}\label{eq:EP}
\Psi_{EP} = \frac{1}{2}\tan^{-1}\left(\frac{\sum_i w_i \sin(2\phi_i)}{\sum_i w_i \cos(2\phi_i)}\right)
\end{equation}
where the index $i$ runs over all charged particles in the event, and the weight $w_i$ is the $p_T$ of particle $i$ \cite{standardEP}.  The method of determining the event plane in hit-based detectors, such as the ZDC-SMD, is similar.  

The event plane resolution in each of the detectors is calculated from the $\eta$-subevent method \cite{standardEP}.  In the 0-50\% centrality range, the resolution of the event plane in the TPC ranges from 0.55 to 0.8 and the resolution of the FTPC event plane is between 0.09 and 0.3.  Since the ZDC-SMD is only sensitive to directed flow, $v_2^{jet}$\{ZDC-SMD\} is measured with respect to the first-order event plane (through the mixed harmonics method \cite{standardEP}).  The resolution of the second-order flow with respect to the first-order event plane in the ZDC-SMD is less than 0.1.  

\section{Results \& Conclusions}

\begin{figure}[tb]
\includegraphics[width=\linewidth,trim = 0mm 0mm 25mm 0mm,clip=true]{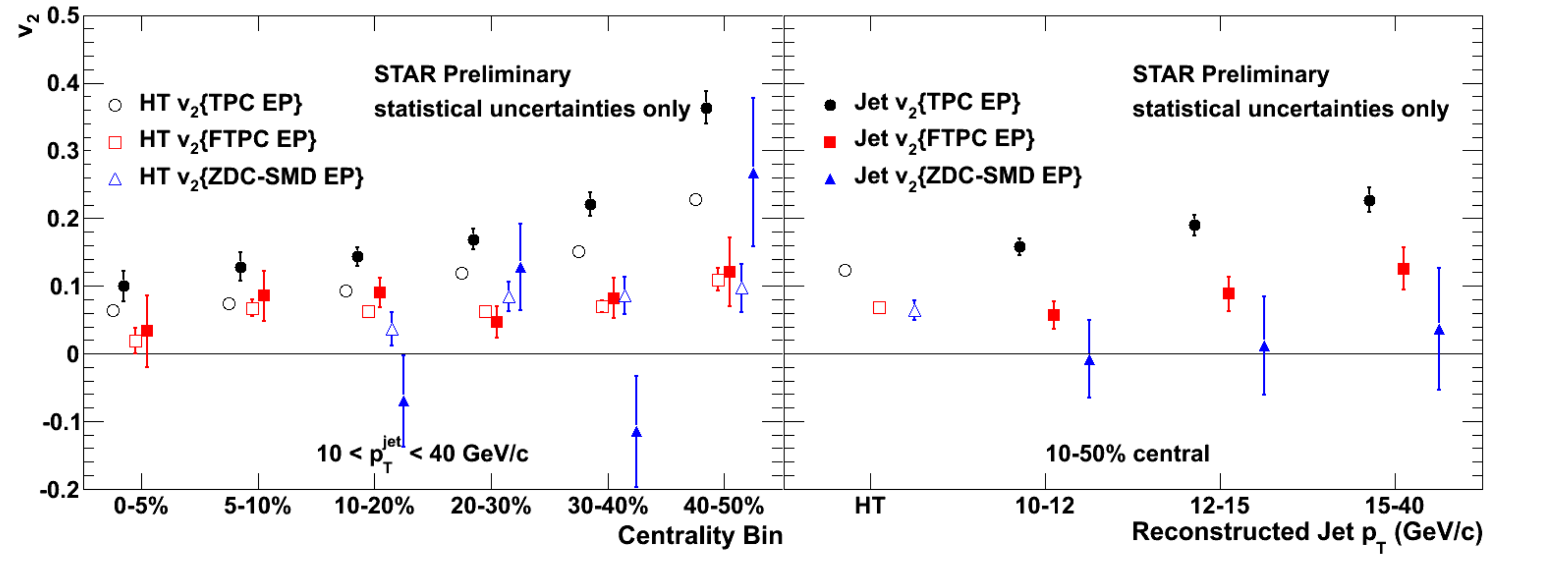}
\caption{\label{fig:v2} Left: Jet $v_2$ (closed symbols) and HT trigger $v_2$ (open symbols) are shown with respect to centrality when the event plane is determined by three sets of detectors: the TPC (black circles), the FTPC (red squares), and the ZDC-SMD (blue triangles).  Right: Jet $v_2$ is shown with respect to reconstructed jet $p_T$.  HT $v_2$ is also shown.  Jets are reconstructed from charged tracks and neutral calorimeter towers with $p_T > 2$ GeV/$c$ and $E_T > 2$ GeV, respectively.  Additionally, the reconstructed jets must contain a neutral tower with $E_T > 5.5$ GeV which fired the online HT trigger.}
\end{figure}

Jet $v_2$ can be investigated differentially with respect to centrality and reconstructed jet $p_T$.  Jet $v_2$ is measured in six centrality bins and compared to the $v_2$ of HT trigger towers ($v_2^{HT}$, with $E_{T} > 5.5$ GeV) in Figure~\ref{fig:v2}.  The artificial jet -- event plane bias is clearly seen in the increase of $v_2^{jet}$\{TPC\} over $v_2^{jet}$\{FTPC\}, due to jet fragments included in the TPC event plane calculation.  In the most peripheral bin (40-50\%), the jet -- event plane bias introduces a 25\% $v_2$ signal.  Additionally, $v_2^{jet}$\{TPC\} is significantly higher than $v_2^{HT}$\{TPC\}, indicating that the jet -- event plane bias is stronger when jets contain additional high-$p_T$ fragments (even though those fragments with $p_{T} > 2$ GeV/$c$ are not included in the event plane calculation).

A comparison of $v_2^{jet}$ and $v_2^{HT}$ can give insights into the biases involved in the jet definition described in Section~\ref{sec:jet}.  It is observed that $v_2^{jet}$\{FTPC\} is consistent with $v_2^{HT}$\{FTPC\}, indicating that the surface bias or bias towards unmodified jets, which is the physical mechanism for jet $v_2$, is largely driven by the HT trigger requirement, and only to a lesser extent by the 2 GeV/$c$ $p_T$ cut.  At this stage, due to limited statistics, we can not draw any conclusions about $v_2^{jet}$\{ZDC-SMD\}, although it is observed that $v_2^{HT}$\{ZDC-SMD\} is non-zero.  The finite values of $v_2^{jet}$\{FTPC\} and $v_2^{HT}$\{ZDC-SMD\} are evidence of the pathlength-dependence of parton energy loss.  

At this point, no centrality dependence is observed within the current statistical precision of this measurement.  Interpretation of this result is further convoluted by the fact that the reconstructed jet energy is dependent on centrality, because more background particles are clustered into the jet cone in central events than in peripheral events.  Future studies will investigate the correspondence between the reconstructed jet $p_T$ and the original parton energy, and the related systematic uncertainties.  

Figure~\ref{fig:v2} shows a slight increase in $v_2^{jet}$\{FTPC\} with reconstructed jet $p_T$.  Furthermore, we observe that $v_2^{jet}$\{FTPC\} is typically larger than $v_2^{jet}$\{ZDC-SMD\}.  In single-particle measurements, this difference can be attributed to flow in the participant plane frame being larger than flow in the reaction plane frame \cite{RPPP}.  This result could therefore be indicative of jet quenching being more sensitive to the participant plane geometry than the reaction plane geometry.  

Measurements of jet $v_2$ are a tool for investigating the pathlength-dependence of parton energy loss in the QGP, as well as the biases involved in jet reconstruction in heavy ion collisions.  We show a first measurement of the $v_2$ of reconstructed jets in STAR, which is indicative of pathlength-dependent parton energy loss in high-$Q^2$ processes in heavy ion collisions at RHIC.

\bibliography{QMbiblio}
\bibliographystyle{elsarticle-num}

\end{document}